\documentclass[twocolumn]{aastex631}
\usepackage{amsmath}

\begin{document}

%\title{Timing analysis of the accreting millisecond pulsar SAX J1808.4-3658 during its 2022 outburst}
\title{Timing analysis of the 2022 outburst of the accreting millisecond X-ray pulsar SAX J1808.4$-$3658: hints of an orbital shrinking}

\author[0000-0003-4795-7072]{Giulia Illiano} \affiliation{INAF-Osservatorio Astronomico di Roma, Via Frascati 33, I-00078, Monteporzio Catone (RM), Italy}
\affiliation{Tor Vergata University of Rome, Via della Ricerca Scientifica 1, I-00133 Roma, Italy}

\author[0000-0001-6289-7413]{Alessandro Papitto}
\affiliation{INAF-Osservatorio Astronomico di Roma, Via Frascati 33, I-00078, Monteporzio Catone (RM), Italy}

\author[0000-0002-0118-2649]{Andrea Sanna}
\affiliation{Dipartimento di Fisica, Università degli Studi di Cagliari, SP Monserrato-Sestu km 0.7, I-09042 Monserrato, Italy}

\author[0000-0002-7252-0991]{Peter Bult}
\affiliation{Department of Astronomy, University of Maryland, College Park, MD 20742, USA}
\affiliation{Astrophysics Science Division, NASA Goddard Space Flight Center, Greenbelt, MD 20771, USA}

\author[0000-0001-7915-996X]{Filippo Ambrosino}
\affiliation{INAF-Osservatorio Astronomico di Roma, Via Frascati 33, I-00078, Monteporzio Catone (RM), Italy}
\affiliation{INAF-Istituto di Astrofisica e Planetologia Spaziali, Via Fosso del Cavaliere 100, I-00133 Rome, Italy}
\affiliation{Sapienza Università di Roma, Piazzale Aldo Moro 5, I-00185 Rome, Italy}

\author[0000-0002-0943-4484]{Arianna Miraval Zanon}
\affiliation{INAF-Osservatorio Astronomico di Roma, Via Frascati 33, I-00078, Monteporzio Catone (RM), Italy}

\author[0000-0001-7611-1581]{Francesco Coti Zelati}
\affiliation{Institute of Space Sciences (ICE, CSIC), Campus UAB, Carrer de Can Magrans s/n, E-08193 Barcelona, Spain} 
\affiliation{Institut d’Estudis Espacials de Catalunya (IEEC), Carrer Gran Capità 2–4, E-08034 Barcelona, Spain} 
\affiliation{INAF, Osservatorio Astronomico di Brera, Via E. Bianchi 46, I-23807, Merate (LC), Italy}

\author[0000-0002-0018-1687]{Luigi Stella}
\affiliation{INAF-Osservatorio Astronomico di Roma, Via Frascati 33, I-00078, Monteporzio Catone (RM), Italy}

\author[0000-0002-3422-0074]{Diego Altamirano}
\affiliation{Physics \& Astronomy, University of Southampton, Southampton, Hampshire SO17 1BJ, UK}

\author[0000-0003-1285-4057]{Maria Cristina Baglio}
\affiliation{Center for Astro, Particle, and Planetary Physics, New York
University Abu Dhabi, P.O. Box 129188, Abu Dhabi, 100190,
UAE}
\affiliation{INAF, Osservatorio Astronomico di Brera, Via E. Bianchi 46, I-23807, Merate (LC), Italy}

\author[0000-0002-8201-1525]{Enrico Bozzo}
\affiliation{ISDC, University of Geneva, Chemin d’Ecogia 16, CH-1290 Versoix, Switzerland}

\author[0000-0001-5458-891X]{Luciano Burderi}
\affiliation{Dipartimento di Fisica, Università degli Studi di Cagliari, SP Monserrato-Sestu km 0.7, I-09042 Monserrato, Italy}

%\author[0000-0001-6278-1576]{Sergio Campana}
%\affiliation{INAF, Osservatorio Astronomico di Brera, Via E. Bianchi 46, I-23807, Merate (LC), Italy}

%\author[0000-0001-8804-8946]{Deepto Chakrabarty}
%\affiliation{MIT Kavli Institute for Astrophysics and Space Research, Massachusetts Institute of Technology, Cambridge, MA 02139, USA}

\author[0000-0002-5069-4202]{Domitilla de Martino}
\affiliation{INAF–Osservatorio Astronomico di Capodimonte, Salita Moiariello 16, I-80131 Napoli, Italy}

\author[0000-0003-0331-3259]{Alessandro Di Marco}
\affiliation{INAF-Istituto di Astrofisica e Planetologia Spaziali, Via Fosso del Cavaliere 100, I-00133 Rome, Italy}

\author[0000-0002-3220-6375]{Tiziana di Salvo}
\affiliation{Università degli Studi di Palermo, Dipartimento di Fisica e Chimica, via Archirafi 36, I-90123 Palermo, Italy}

\author[0000-0003-1429-1059]{Carlo Ferrigno}
\affiliation{ISDC, University of Geneva, Chemin d’Ecogia 16, CH-1290 Versoix, Switzerland}

\author[0000-0001-6894-871X]{Vladislav Loktev}
\affiliation{Department of Physics and Astronomy, FI-20014 University of Turku, Finland}

\author[0000-0001-5674-4664]{Alessio Marino}
\affiliation{Institute of Space Sciences (ICE, CSIC), Campus UAB, Carrer de Can Magrans s/n, E-08193 Barcelona, Spain} 
\affiliation{Institut d’Estudis Espacials de Catalunya (IEEC), Carrer Gran Capità 2–4, E-08034 Barcelona, Spain} 
\affiliation{INAF/IASF Palermo, via Ugo La Malfa 153, I-90146 Palermo, Italy}

\author[0000-0002-0940-6563]{Mason Ng}
\affiliation{MIT Kavli Institute for Astrophysics and Space Research, Massachusetts Institute of Technology, Cambridge, MA 02139, USA}

\author[0000-0001-7397-8091]{Maura Pilia}
\affiliation{INAF—Osservatorio Astronomico di Cagliari—via della Scienza 5—I-09047, Selargius, Italy}

\author[0000-0002-0983-0049]{Juri Poutanen}
\affiliation{Department of Physics and Astronomy, FI-20014 University of Turku, Finland}

\author[0000-0001-6356-125X]{Tuomo Salmi}
\affiliation{Anton Pannekoek Institute for Astronomy, University of Amsterdam, Science Park 904, 1090GE Amsterdam, the Netherlands}



\begin{abstract}
We present a pulse timing analysis of \textrm{NICER} observations of the accreting millisecond X-ray pulsar SAX J1808.4$-$3658 during the outburst that started on 2022 August 19.
Similar to previous outbursts, after decaying from a peak luminosity of $\simeq 1\times10^{36} \, \mathrm{erg \, s^{-1}}$ in about a week, the pulsar entered in a $\sim 1$ month-long reflaring stage.  
Comparison of the average pulsar spin frequency during the outburst with those previously measured confirmed the long-term spin derivative of $\dot{\nu}_{\textrm{SD}}=-(1.15\pm0.06)\times 10^{-15}$~Hz\,s$^{-1}$, compatible with the spin-down torque of a $\approx 10^{26}$~G~cm$^3$ rotating magnetic dipole. 
For the first time in the last twenty years, the orbital phase evolution shows evidence for a decrease of the orbital period. The long-term behaviour of the orbit is dominated by a $\sim 11$~s modulation of the orbital phase epoch consistent with a $\sim 21$~yr period. 
We discuss the observed evolution in terms of a coupling between the orbit and variations in the mass quadrupole of the companion star. 
\end{abstract}

\keywords{accretion, accretion discs -- pulsars: individual (SAX J1808.4$-$3658)  -- stars: neutron -- X-rays: binaries
}


\section{Introduction}
The transient low-mass X-ray binary SAX J1808.4$-$3658 (hereafter SAX J1808) was discovered in 1996 with the X-ray satellite \textrm{BeppoSAX} during an X-ray outburst \citep{Discovery_1996}. 
Three type-I X-ray bursts were detected \citep{intZand_2001}, permitting to identify the accretor as a neutron star (NS). The distance was later estimated by \citet{Galloway_2006} to be $\sim 3.5$ kpc.
The detection of 401 Hz X-ray pulsations with \textrm{RXTE} during the following outburst in 1998 marked the discovery of the first accreting millisecond X-ray pulsar (AMXP; \citealt{Wijnands_1998}). Timing analysis revealed that the NS is in an orbit with a $\approx 0.05 \, {M_\odot}$ brown dwarf companion \citep{Bildsten_2001} with a 2.01 hr orbital period \citep{Chakrabarty_Morgan_1998}. Since its discovery, the source has undergone ten $\sim$1 month-long outbursts with $\sim$2--3 years recurrence. This makes it the AMXP that has shown the largest number of outbursts of sufficient duration for in-depth investigation of its long-term timing properties \citep{Marino_2019, DiSalvo_2022}. It is thus the most thoroughly studied AMXP.
The X-ray luminosity typically reaches $L_{\textrm{X}} \sim \mathrm{few} \times 10^{36} \, \mathrm{erg \, s^{-1}}$ \citep{Gilfanov_1998} at the peak of the outbursts, and decreases down to $L_{\textrm{X}} \sim \mathrm{few} \times 10^{31} \, \mathrm{erg \, s^{-1}}$ during quiescence \citep{Stella_2000, Campana_2004}. Coherent 401 Hz X-ray pulsations are observed only during outbursts and interpreted in terms of magnetic channeling of the in-flowing matter onto the NS magnetic poles. During the 2019 outburst, \citet{Ambrosino_2021} discovered coherent ms optical and UV pulsations. The bright pulsed luminosity ($L_{\textrm{optical}} \approx 3 \times 10^{31} \, \mathrm{erg \, s^{-1}}$, and $L_{\textrm{UV}} \approx 2 \times 10^{32} \, \mathrm{erg \, s^{-1}}$) seen at those wavelengths challenged the expectations of the standard accretion models.

On 2022 August 19, the MAXI/GSC nova alert system \citep{Atel_MAXI_2022} detected the occurrence of a new outburst of SAX J1808, later confirmed by rapid targeted follow-up \textrm{NICER} observations \citep{Atel_Sanna_2022}.
Here, we report on the high-cadence monitoring campaign performed with the \textrm{NICER} guest observer program ID 5574 (PI: A.~Papitto). We focus on the pulse phase timing analysis carried out on the X-ray pulsations detected throughout the outburst to measure the pulsar spin frequency and the binary orbital ephemeris. These values are compared with those observed during previous outbursts \citep{Hartman_2008, Hartman_2009, DiSalvo_2008, Burderi_2009, Patruno_2016, Sanna_2017, Bult_2020} to derive the long-term spin and orbital evolution of the pulsar and discuss the implications for the models of AMXPs.

\section{Observations} \label{sec:obs}
The NASA X-ray telescope Neutron star Interior Composition Explorer (\textrm{NICER}) \citep{NICER_Gendreau_2012} monitored SAX J1808 from 2022 August 19 (MJD 59810) until 2022 October 31 (MJD 59883) (ObsIDs starting with 505026 and 557401).
The top panel of Fig.~\ref{Fig:subplots} shows the 0.5--10~keV  light curve. Visibility constraints prevented \textrm{NICER} from obtaining a homogeneous coverage of the outburst. 
The data were reduced and processed using \texttt{HEASoft} version 6.30 and \texttt{nicerl2} task (\texttt{NICERDAS} version 7a), retaining events in the 0.5--10~keV energy range.
We corrected the photon arrival times to the Solar System Barycenter (SSB) using the JPL ephemerides DE405 \citep{Standish_DE405}. We adopted the source coordinates R.A. (J2000)=$18^{\mathrm{h}}08^{\mathrm{m}}27\fs647(7)$ and DEC. (J2000)=$-36\degr58'43\farcs90(25)$ from \citet{Bult_2020}. We estimated the background contributions to our data with the \texttt{nibackgen3C50} tool \citep{Remillard_2022}.\\

The \textrm{NICER} monitoring started when SAX J1808 had almost attained its peak count rate of $\sim 300 \, \mathrm{c\,s^{-1}}$ (top panel of Fig.~\ref{Fig:subplots}).  

To estimate the peak luminosity, we extracted the spectrum collected in the observation on 2022 August 24 (ObsId 5574010102), and modelled it within the \texttt{XSPEC} spectral fitting package \citep{Arnaud_1996}. We accounted for absorption effects using the \textsf{tbabs} model with \textsf{wilm} abundances \citep{Wilms_2000} and \textsf{vern} cross-sections \citep{Verner_1996}. We described the continuum emission using a combination of a disc blackbody (\textsf{diskbb}) and a Comptonization component (\textsf{nthComp}), adding three Gaussian emission lines. The electron temperature was held fixed to 30 keV in the fit. We obtained a satisfactory fit ($\chi^2$/dof=865.70/850).
We then calculated the 0.6--10~keV X-ray unabsorbed flux using the convolution model \textsf{cflux}. The corresponding peak luminosity is $\approx 1 \times 10^{36} \, \mathrm{erg \, s^{-1}}$ (assuming a distance of $3.5 \, \mathrm{kpc}$; \citealt{Galloway_2006})

After $\sim$ 5 days from the first detection, the decay phase begun, until the source entered its typical reflaring stage \citep{Cui_1998, Wijnands_2001, Hartman_2008, Patruno_2021}, which was observed with \textrm{NICER} for more than a month \citep{Illiano_2022ATel}.
The light curve of the 2022 outburst slightly differed from the typical profile shown by SAX J1808. The usually short-lived peak exhibited the longest duration observed so far, and the slow decay/rapid drop lasted much less ($\sim$ 10--15\,d) than usual. No type-I X-ray bursts were detected during \textrm{NICER} observations, unlike most other outbursts \citep[see e.g.][]{intZand_2001, Chakrabarty_2003, Galloway_2006, Bult_2015, Bult_2020}.

\section{Results}
\subsection{Coherent timing} 
\begin{figure}
   \centering
   \includegraphics[width=8.5cm]{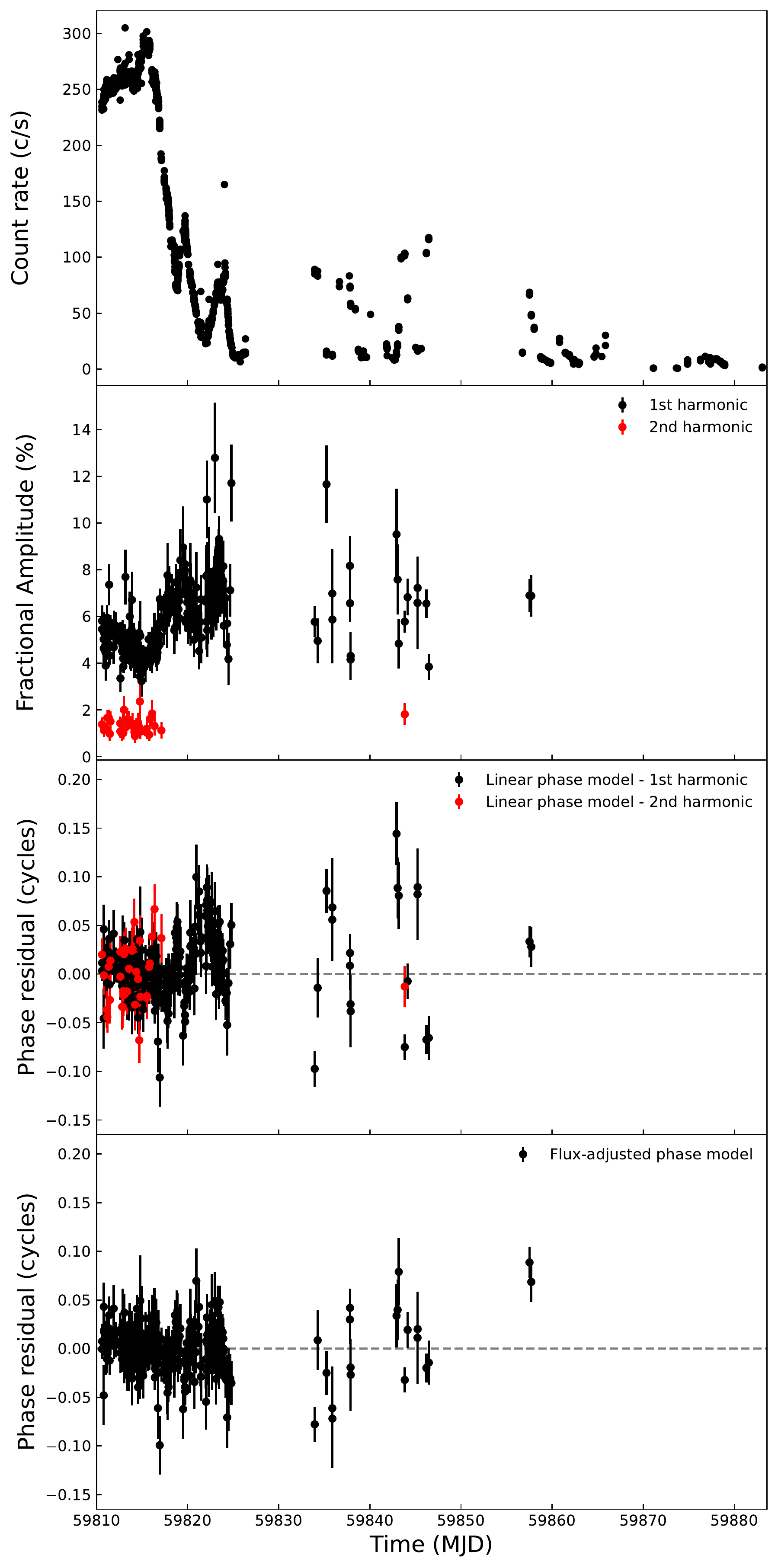}
   \caption{Temporal evolution of the 2022 outburst monitored with \textrm{NICER}. Top panel: the 0.5--10~keV light curve using 200-s bins. Second panel: pulse fractional amplitude for the first harmonic (black) and the second harmonic (red). %Third panel: pulse phase for the first harmonic (black) and the second component (red). 
   Third and fourth panels: phase residuals relative to the linear phase model for the first harmonic (black) and the second harmonic (red), and flux-adjusted phase models, respectively (also see Table \ref{table:Timing_sol}). The phase residuals relative to the quadratic model are not plotted as they are similar to those of the linear model (see text). 
   } \label{Fig:subplots}
\end{figure}

In order to correct the photon arrival times for the pulsar orbital motion in the binary system, we first performed a preliminary search on the epoch of passage at the ascending node, $T_{\textrm{asc}}$, exploiting the variance of the epoch-folding search as a statistical estimator. We fixed the orbital period and the projected semi-major axis equal to the values found in the timing solution of the 2019 outburst \citep{Bult_2020}, and we used the best $T_{\textrm{asc}}$ found as a starting point for the pulse phase timing.
After correcting the photon arrival times with this preliminary orbital solution, we divided our data set into 1000-s long segments and folded them around our best estimate of the spin frequency $\nu_{\textrm{F}}$ using 16 phase bins. We modelled our pulse profiles with a constant plus two harmonic components, retaining only data in which the signal was detected with an amplitude significant at more than a 3$\sigma$ confidence level.
Throughout the outburst, the amplitude of the fundamental (black dots in the second panel of Fig.~\ref{Fig:subplots}) is higher than the second harmonic (red dots in the same panel), increasing by approximately $\sim$1--2 percentage points when the rapid drop phase of the outburst took place and slightly again at the beginning of the flaring tail.

We modelled the time evolution of the phase of the fundamental, using
\citep[see e.g..][]{Burderi_2007, Papitto_2007, Sanna_2022}:
\begin{equation} \label{eq:phase_residuals}
    \Delta \phi (t) = \phi_0 - \Delta \nu (t-T_0) - \frac{1}{2} \, \dot{\nu} \, (t-T_0)^2 + R_{\textrm{orb}}(t).
\end{equation}
Here, $\nu$ is the pulsar spin frequency, $T_0$ is the chosen reference epoch, $\phi_0$ is the pulse phase at $T_0$, $\Delta \nu=\nu(T_0)-\nu_{\textrm{F}}$, while $R_{\textrm{orb}}(t)$ is the residual Doppler modulation due to a difference between the adopted orbital parameters and the actual ones \citep[see e.g.][]{Deeter_Boyton_Pravdo_1981}.
Table \ref{table:Timing_sol} shows the best-fitting orbital and spin parameters we obtained. To take into account the large value of the reduced $\chi^2$ obtained from the fit, we rescaled the uncertainties of the fit parameters by the square root of that value \citep[see e.g.,][]{Finger_1999}.
We estimated the systematic uncertainty on the spin frequency due to the positional uncertainties of
the source using the expression $\sigma_{\nu_{\textrm{pos}}} \leqslant \nu_0 \, y \, \sigma_\gamma \, (1+\sin^2{\beta})^{1/2} \, 2 \pi/P_{\oplus}$,
%, and $\sigma_{\dot{\nu}_{\textrm{pos}}} \leqslant \nu_0 \, y \, sigma_\gamma \, (1+\sin^{\beta})^{1/2} \, (2 \pi/P_{\oplus})^2$, 
where $y = r_{E}/c$ is the semi-major axis of the Earth orbit in light seconds, $\sigma_\gamma$ is the positional error circle, $\beta$ is the source latitude in ecliptic coordinates, and $P_{\oplus}$ is the Earth orbital period \citep[see e.g.,][]{Lyne_GrahamSmith_1990, Burderi_2007, Sanna_2017}. Adopting the positional uncertainties reported by \citet{Bult_2020}, we estimate $\sigma_{\nu_{\textrm{pos}}} \leqslant 5 \times 10^{-8} \, \mathrm{Hz}$. We added in quadrature this systematic uncertainty to the statistical error of the spin frequency reported in Table \ref{table:Timing_sol}.

We base the discussion of the phase evolution during the 2022 outburst on the properties of the fundamental frequency. In fact, below 3~keV, the second harmonic was often too weak to be detected by \textrm{NICER} \citep{Patruno_2009,Bult_2020}. We modelled the phase delays using either a constant frequency model (i.e., setting $\dot{\nu}=0$ in Eq. \eqref{eq:phase_residuals}; see Table \ref{table:Timing_sol}) or a constant spin frequency derivative. The quadratic fit returns a value of the average frequency derivative of $\dot{\nu}$ = $2.4(4.0) \times10^{-15}$\,Hz\,s$^{-1}$ ($\chi ^2$/dof $=698.2/284$) which is compatible with zero. The probability of a chance improvement of the $\chi^2$ compared to the constant frequency model obtained with a F-test is $\sim 0.5$, indicating that the addition of such a component does not produce a significant improvement in the data description.

A strong variability of the phase and shape of the pulse profiles characterised all SAX J1808 outbursts observed so far \citep[see the reviews by][]{Patruno_2021,DiSalvo_2022}. This strongly limited the ability to measure the NS spin evolution during individual outbursts from pulse phase timing. Pulse phases measured from the second harmonic generally showed a more regular behaviour compared to the fundamental. \citet{Burderi_2006} exploited this property to infer a spin-up rate of $\dot{\nu}=4.4(8)\times10^{-13}$\,Hz\,s$^{-1}$ during the 2002 outburst. Such a value is only slightly larger than that  expected considering the material torque exerted by accretion through a Keplerian disc in-flow truncated a few tens of km from the NS ($\simeq 2\times 10^{-13}$\,Hz\,s$^{-1}$  for a $1.4 \, {M_\odot}$ NS accreting at a rate of $10^{-9} \,{M_\odot}$\,yr$^{-1}$ from a disc truncated at 20 km from the NS; see e.g. \citealt[and references therein]{DiSalvo_2019}). \citet{Hartman_2008,Hartman_2009} attributed instead much of the observed phase variability to a red noise process affecting the pulse phases on timescales similar to the outburst duration; this led to tighter upper limits on the spin frequency derivative ($|\dot{\nu}|<2.5\times10^{-14}$\,Hz\,s$^{-1}$). \citet{Patruno_2009} characterised such a noise process in terms of a correlation between the pulse phase and the X-ray flux. Azimuthal drifts of the hot spot location on the NS surface related to a movement of the inner disc truncation radius at different mass accretion rates could explain such a phase-flux correlation. However, a broadly different correlation characterised each of the outbursts of SAX J1808.
In this context, \citet{Bult_2020} found the best description of the evolution of the pulse phases measured by \textrm{NICER} in 2019 by using a phase-flux correlation term related to hot spots drifts, $R_{\textrm{flux}}(t)=b F_X(t)^{\Gamma}$, where $F_X$ is the X-ray bolometric flux, $b=-0.87(3)$, and $\Gamma=-0.2$ fixed. These values were broadly consistent with those expected according to numerical simulations of accretion onto a fast-rotating NS. The fixed power law index arises from the linear scaling of the azimuthal position of the hot spot with the magnetospheric radius, which was recently predicted to depend on the mass accretion rate as $\dot{M}^{-1/5}$ \citep{Kulkarni_2013}.

Because of the large phase residuals with respect to the linear model, following \citet{Bult_2020} we also attempted to replace in Eq.~\eqref{eq:phase_residuals} the spin frequency derivative term with a component including a dependence of the pulse phase on the flux, here considered to be traced by the 0.5--10~keV count rate $R(t)$ ($R_{\textrm{flux}}(t)=b R(t)^{\Gamma}$). 
The resulting $\chi^2$ shows a significant improvement with respect to the linear phase model (F-test probability of $\sim 8.5 \times 10^{-28}$; see also panel 4 in Fig.~\ref{Fig:subplots}).

If we restrict our analysis to the first $\sim 8$ days of the outburst, i.e., until the source faded to roughly a fifth of the peak flux, then the addition of either a spin frequency derivative or of a phase-flux correlation component did not improve the phase description compared to a constant spin frequency model (F-test probability of the quadratic model with respect to the linear one of $\sim 0.7$; F-test probability of the flux-adjusted model with respect to the linear one of $\sim 0.8$). The phase behaviour is compatible with a constant spin frequency, with a $90\%$ c.l. upper limit on the spin frequency derivative of $1.9 \times 10^{-13}$\,Hz\,s$^{-1}$ (same order of magnitude as the expected one for the accretion-driven spin-up, discussed above).

This points to an anti-correlation between the phase delays and the source flux, observed in Fig. \ref{Fig:subplots}, holding only for count rates lower than $\sim 100$\,c\,s$^{-1}$, i.e., in the reflaring phase. Even though we lacked a coverage of the rising part of the outburst, i.e., where most of the flux dependence of the phases was present in 2019 data  \citep[see Fig.~1 in][]{Bult_2020}, we found an even more pronounced phase-flux anti-correlation than in the previous outburst. 
Since the $\Gamma$ index we obtained is not consistent with the hot spots drifts predicted by numerical simulations of accreting pulsars \citep{Kulkarni_2013}, the phase shifts are not driven by the changing size of the magnetosphere, but are instead inversely proportional to the mass accretion rate (similar to the case of the AMXP MAXI J1816$-$195; \citealt{Bult_2022}).
On the other hand, no such variation was seen when the flux varied by a three-times larger factor during the peak and the decay phase. The steep index of the phase-flux correlation we found ($\delta \phi \sim 1/F_X$) naturally explains why introducing this term determines a significant improvement of the quality of the residuals of the fit performed on the whole dataset, even though phase fluctuations are essentially observed only at low count rates.\\

\begin{table} 
\renewcommand{\arraystretch}{1.2}
\centering
\caption{Timing solution for SAX J1808 2022 outburst.}             
\label{table:Timing_sol}      
\begin{tabular}{l c}          
\hline\hline                       
Parameter & Value\\
\hline
Epoch (MJD) & 59810.5956860\\
$a_1 \sin{i}$ (lt-s) & 0.0628033(57)\\
$P_{\textrm{orb}}$ (s) & 7249.1600(13)\\
$T_{\textrm{asc}}$ (MJD) & 59810.6179996(17)\\
\hline
Linear phase model\\
\hline
$\nu$ (Hz) & 400.975209557(50)\\
$\chi ^2$/dof & 699.1/285\\
\hline
Flux-adjusted phase model\\
\hline
$\nu$ (Hz) & 400.975209535(50)\\
b & $1.44(49)$\\
$\Gamma$ & $-0.81(12)$\\
$\chi ^2$/dof & 450.0/283\\
\hline
\end{tabular}
\tablecomments{The timing solution was obtained adopting the source coordinates from \citet{Bult_2020}. Uncertainties are the $1\sigma$ statistical errors.}
\end{table}
\noindent

\subsection{Long-term spin frequency evolution}
\begin{figure}
   \centering
   \includegraphics[width=8.5cm]{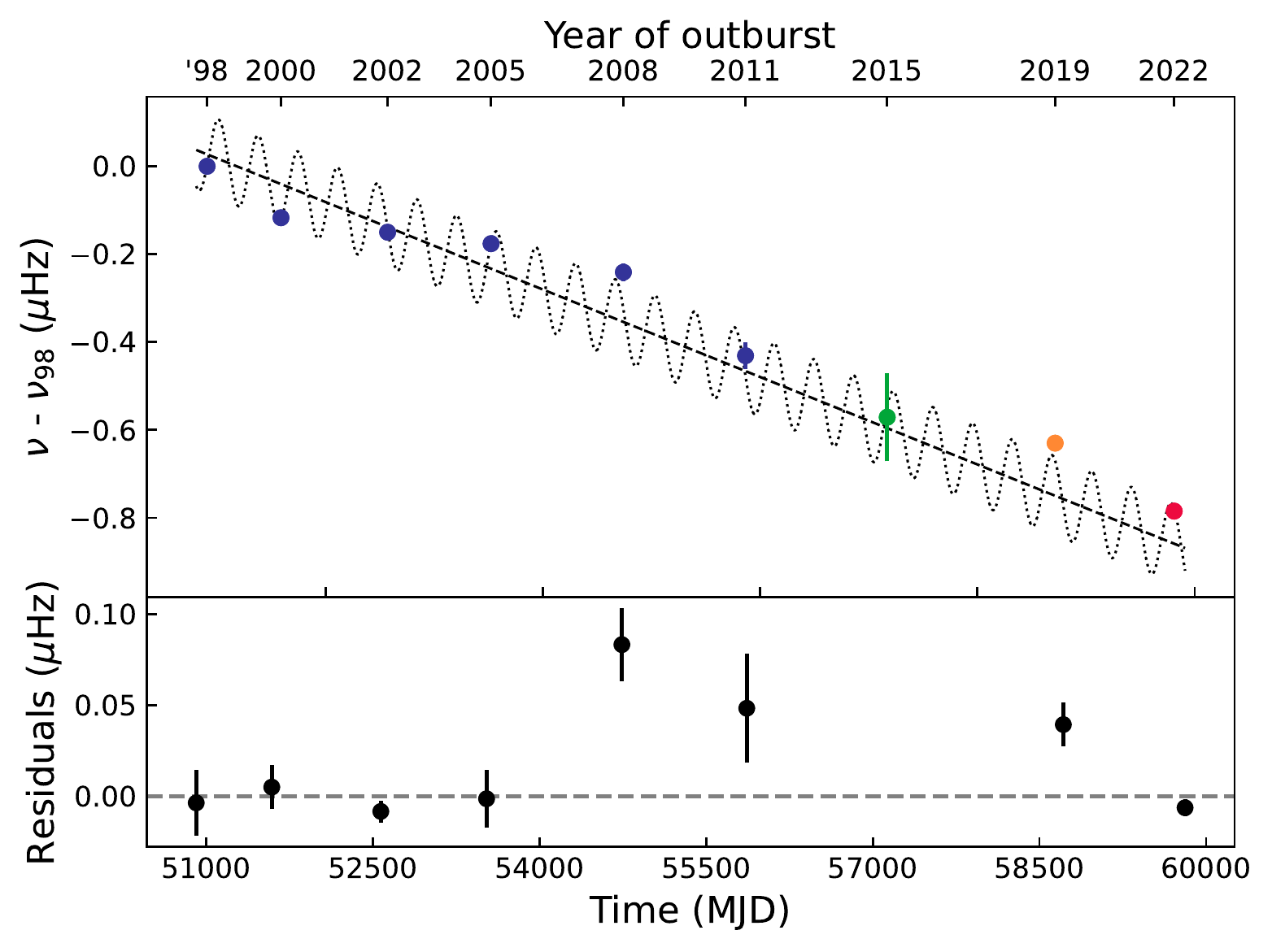}
   \caption{Top panel: spin frequency evolution of SAX J1808 since the 1998 outburst. Blue points show measurements made with RXTE from the 1998 outburst to that of 2011 \citep{Hartman_2008, Hartman_2009, Atel_Papitto_2011}, the green dot represents the best estimate for the 2015 outburst \citep{Sanna_2017}, and the orange one is from the linear model of the 2019 outburst \citep{Bult_2020}. The red dot is from this work, having corrected the data with the source coordinates from \citet{Hartman_2008} and fitted the phase delays with a linear model. All frequencies are expressed relative to the 1998 spin frequency, $\nu_{98}=400.975210371$ Hz \citep{Hartman_2008}.
   The dotted line indicates the best-fitting function including the Doppler modulation due to source coordinates error, and the dashed line is the corresponding linear function. Bottom panel: residuals relative to the best-fitting function. We did not include the 2015 spin frequency estimate because its uncertainty is about a factor twenty larger than the others and compatible with the amplitude of Doppler modulation.} \label{Fig:freq}
\end{figure}

The ten SAX J1808 outbursts observed so far, the most numerous for any AMXP, enable a detailed study of the long-term spin frequency evolution through a comparison of the measurements obtained in each of the outbursts. Previous works \citep[see e.g.][]{Patruno_2012, Sanna_2017, Bult_2020} found that the spin frequency decreased at an average rate of $\dot{\nu}_{\textrm{SD}} \simeq -10^{-15}$\,Hz\,s$^{-1}$, compatible with the energy losses expected from a $\approx 10^{26}$\,G\,cm$^3$ rotating magnetic dipole.
\citet{Bult_2020} also found that the spin frequencies measured  by correcting the pulse arrival times with the position measured by \citet{Hartman_2008} showed a yearly modulation due to an offset of $\delta\lambda=(0.33\pm0.10)\arcsec$ and $\delta\beta=(-0.60\pm0.25)\arcsec$ in the assumed Galactic longitude and latitude of the pulsar, respectively. In order to compare the frequency observed in the 2022 outburst with the past values, in this part of the analysis we corrected the photon arrival times to the SSB adopting the optical coordinates from \citet{Hartman_2008}, in analogy with previous works. Using a linear phase model, we obtained $\nu=400.9752095863(45)$~Hz, higher than $\sim 8 \times 10^{-7}$ Hz compared to the values obtained with the coordinates from \citet{Bult_2020}. We then modelled the long-term frequency evolution (see Fig.~\ref{Fig:freq}) with a function including a constant spin-down and a position correction term:
\begin{equation}
    \Delta \nu (t) = \delta \nu _{98} + \dot{\nu}_{\textrm{SD}} (t-T_{98}) + \delta \nu_{\textrm{pos}}(t, \lambda, \beta).
\end{equation}
Here, $\delta \nu _{98} = \nu(t) - \nu_{98}$ is the spin frequency difference compared to the 1998 value, $\nu_{98}=400.975210371$ Hz \citep{Hartman_2008}, $T_{98}=50914.8$ MJD \citep{Hartman_2008}, and $\delta \nu_{\textrm{pos}}(t, \lambda, \beta)$ is the Doppler correction \citep[see, e.g.,][]{Bult_2020}. We found $\delta \nu_{98} = 2.7(1.9) \times 10^{-8} \, \mathrm{Hz}$,
$\dot{\nu}_{\textrm{SD}} = -1.152(56) \times 10^{-15}$\,Hz\,$s^{-1}$, $\delta \lambda = 0\farcs42(15)$, and $\delta \beta = -0\farcs93(38)$, with $\chi ^2$/dof $ = 34.9/5$. 
Uncertainties of our best-fitting values were estimated from the parameters' range required to increase the $\chi^2$ from the fit by an amount $\Delta \chi^2(\mathrm{C.L.} = 68 \%, p=4) = 4.7$, where $p$ is the number of interesting free parameters \citep{Lampton_Margon_Bowyer, Avni_1976, Yaqoob_1998}. The spin-down trend observed across the previous outbursts is therefore confirmed. Also, the coordinate offsets are compatible within $1\sigma$ with what was found by \citet{Bult_2020}, and correspond to R.A.(J2000) = $18^{\mathrm{h}}08^{\mathrm{m}}27\fs656(12)$, DEC. (J2000)= $-36\degr58'44\farcs222(89)$.

\subsection{Orbital period evolution}
\begin{figure}
   \centering
   \includegraphics[width=8.5cm]{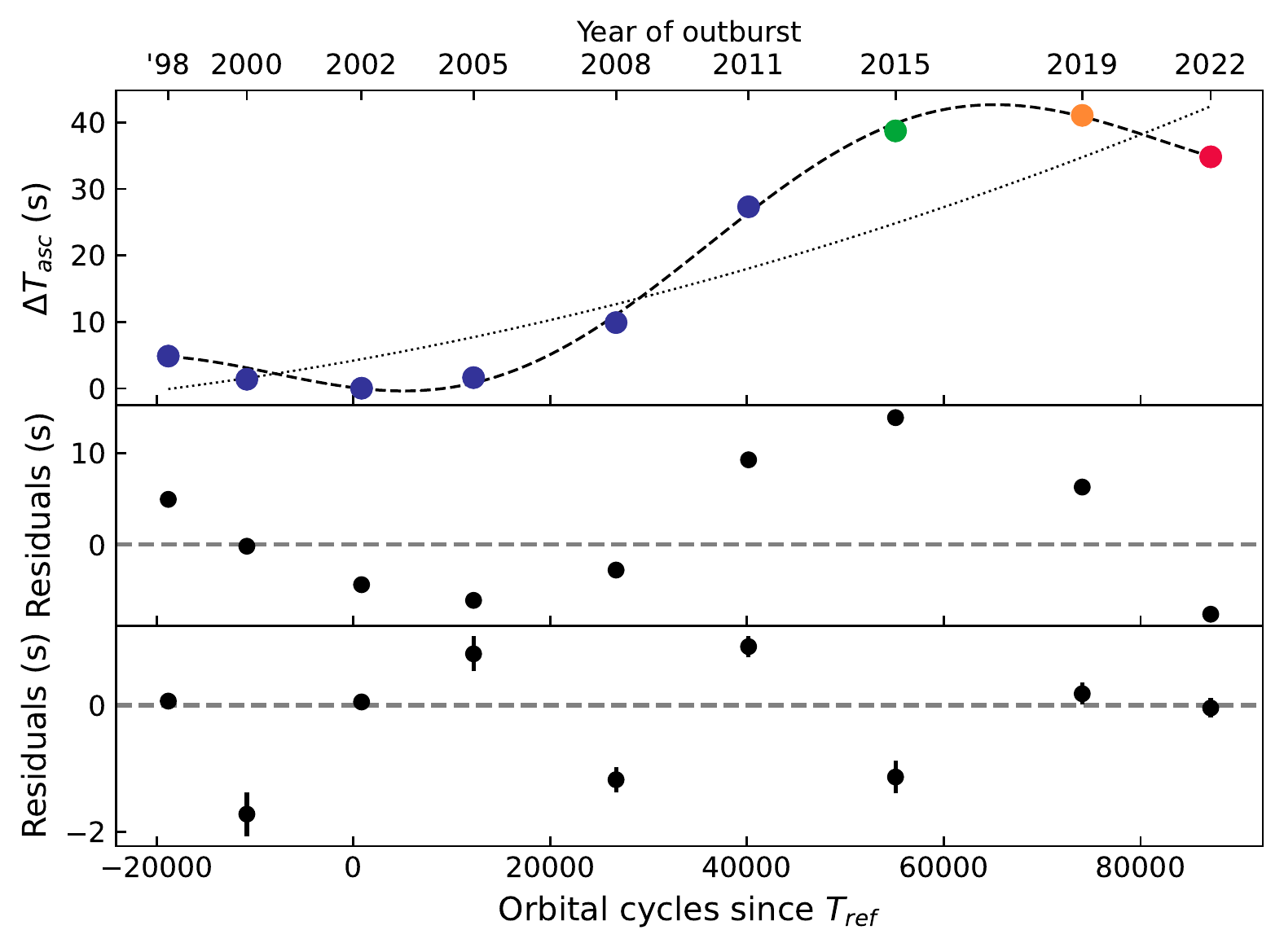}
   \caption{Top panel: long-term evolution of $T_{\textrm{asc}}$ as a function of the number of orbital cycles since the epoch of the 2002 outburst, $T_{\textrm{ref}}=\mathrm{MJD}\, 52499.9602472$. Blue points represent the measurements made with RXTE from the 1998 outburst to that of 2011 \citep{Hartman_2008, Burderi_2009, Atel_Papitto_2011}, the green dot is the best value found for the 2015 outburst \citep{Sanna_2017}, the orange one for the 2019 outburst \citep{Bult_2020}, and the red one is from this work. The dotted line indicates a quadratic fitting function, while the dashed line indicates the best-ﬁtting quadratic$+$sinusoidal function. Middle panel: residuals relative to the quadratic model ﬁt. Bottom panel: residuals relative to the quadratic$+$sinusoidal fitting function.
   We point out that different y-axes are used for the second and the third panels.
   } \label{Fig:Tasc}
\end{figure}

To investigate the orbital evolution, we computed the difference $\Delta T_{\textrm{asc}}$ between the measurements of the epoch of passage at the ascending node during the various outbursts and the values extrapolated from the epoch of passage at the ascending node estimated in the 2002 outburst ($T_{\textrm{ref}}=52499.9602472 \,\mathrm{MJD}$), assuming a constant orbital period ($P_{\textrm{ref}}=7249.156980(4)$ s; \citealt{Hartman_2009}), $\Delta T_{\textrm{asc},i} = T_{\textrm{asc},i} - (T_{\textrm{ref}} + N_{\textrm{orb}} \, P_{\textrm{ref}})$. Here, $T_{\textrm{asc},i}$ is the epoch of passage at the ascending node for the $i$-th outburst, and $N_{\textrm{orb}}$ is the nearest integer number of orbital cycles since $T_{\textrm{ref}}$. 
Until the 2008 outburst, the orbital phase evolution was consistent with an expansion at an average rate of $\simeq 4 \times 10^{-12} \, \mathrm{s \, s^{-1}}$ \citep{Hartman_2008, DiSalvo_2008, Hartman_2009, Burderi_2009}. Subsequent outbursts first suggested an acceleration of the expansion \citep{Patruno_2012}, then a transition to a slower evolution \citep{Sanna_2017, Patruno_2017, Bult_2020}. 
The orbital phase we measured in 2022 data (see the red point in the top panel of  Fig.~\ref{Fig:Tasc}) indicates the first decrease of the orbital period seen from SAX J1808 in the last twenty years. 
Indeed, modelling the $\Delta T_{\textrm{asc}}$ evolution with a constant period derivative (dotted line in Fig. \ref{Fig:Tasc}), leaves evident residuals with a sinusoidal shape ($\chi ^2$/dof$=15579.0/6$, see the middle panel of Fig.~\ref{Fig:Tasc}). 
We then added a sinusoidal term to the relation used to fit the orbital phases:
\begin{multline} \label{eq:fit_Dtasc}
 \Delta T_{\textrm{asc}}(N_{\textrm{orb}}) = \delta T_{\textrm{ref}} + \delta P_{\textrm{ref}} \, N_{\textrm{orb}} + \\
  + \frac{1}{2}  \, \dot{P}_{\textrm{orb}} \, P_{\textrm{ref}} \, N_{\textrm{orb}}^2 + A \, \sin{\Biggr[\frac{2 \, \pi}{P} \, (N_{\textrm{orb}}-N)\Biggr]}.
\end{multline}
Here, $\delta T_{\textrm{ref}}$ is the offset from the 2002 epoch of passage at the ascending node, $\delta P_{\textrm{ref}}$ is the correction to the orbital period at the epoch of the 2002 outburst,  $\dot{P}_{\textrm{orb}}$ is the orbital period derivative, and $A$, $P$ and $N$ are the amplitude, period and phase of the sinusoidal function, respectively. The addition of the last term in Eq.~\eqref{eq:fit_Dtasc} led to a decrease of the fit's $\chi ^2$/dof down to $117.9/3$.
Although statistically speaking the fit is clearly still unacceptable, a F-test indicates that the probability that the improvement occurs by chance is 0.1\%.
The best-fit values are the following: $\delta P_{\textrm{ref}} = 4.63(16) \times 10^{-4} \, \mathrm{s}$, and $\dot{P}_{\textrm{orb}} = -2.82(69) \times 10^{-13} \, \mathrm{s \, s^{-1}}$, $A=11.30(33) \, \mathrm{s}$, and $P = 7.57(21) \times 10^{3} \, \mathrm{d}$. We evaluated the uncertainties by varying the parameters as to obtain a $\Delta \chi^2(\mathrm{C.L.} = 68 \%, p=4) = 4.7$. The amplitude and period of the long-term modulation we found are similar to the values measured by \citet{Sanna_2017} from an analysis of the outbursts observed until 2015.
The large $\chi^2$ of the fit suggests caution in interpreting these results. SAX J1808 orbital variability is similar to that observed in black widow and redback millisecond pulsars, rotation-powered pulsars in close binary orbits that ablate matter from their very low-mass ($\lesssim 1 \,{M_\odot})$ companion star. Yet the presence of a sinusoidal-like modulation of the orbital phase and of a much lower, formally negative, orbital period derivative evolution than previously estimated appear to be solid enough conclusions to draw. 
The sinusoidal modulation is hardly explained through the presence of a third body. The mass function \citep[see, e.g.,][]{Bildsten_2001} of a putative third body would be $\simeq 2.7 \times 10^{-8} \, {M_\odot}$.
Considering a NS mass of $\sim 1.4\, {M_\odot}$ and neglecting the companion mass ($\simeq 0.05\,{M_\odot}$), the implied mass for the hypothetical third body would be $\sim 0.004 \,{M_\odot}$, for a third body inclination similar to the one of the system, $i \sim 69^{\circ}$ \citep{Goodwin_2019}. However, assuming that the orbit of SAX J1808 and of the putative third body is planar, the expected Doppler modulation of the pulsar frequency is $\delta \nu \sim (2 \pi/P) \, A \, \nu  \sim 42 \, \mathrm{\mu Hz}$, which is about two orders of magnitude higher than observed (see Fig. \ref{Fig:freq}).\\ 
\noindent

\section{Discussion} \label{sec:discussion}
The long-term orbital evolution of SAX J1808 has been discussed by several authors \citep[see, e.g.,][]{DiSalvo_2008, Hartman_2008, Hartman_2009, Burderi_2009, Patruno_2012, Sanna_2017, Patruno_2017}. A conservative mass transfer was soon excluded as the mass accretion rate implied by the large  $\dot{P}_{\textrm{orb}}\simeq 4 \times 10^{-12} \, \mathrm{s \, s^{-1}}$, indicated by the first outbursts, is two orders of magnitude larger than $\approx$ $10^{-12}\,{M_\odot}$\,yr$^{-1}$ estimated from the average X-ray flux observed summing outbursts and quiescence periods \citep{Marino_2019}.
\citet{DiSalvo_2008} and \citet{Burderi_2009} discussed the surprisingly large value of $\dot{P}_{\textrm{orb}}$ of SAX J1808 in terms of mass lost by the system at a rate of $\approx 10^{-9}{M_\odot}$\,yr$^{-1}$  from the inner Lagrangian point, e.g., due to irradiation by a rotation-powered pulsar active in quiescence \citep{Burderi_2003}. As also noted by \citet{Hartman_2008} and \citet{DiSalvo_2008}, the fast orbital evolution of SAX J1808 is reminiscent of black widow and redback pulsars. In these systems, the orbital period may change unpredictably with time, with $T_{\textrm{asc}}$ variations ranging from a few seconds to a few tens of seconds over a timescale of tens of years \citep[see, e.g.,][]{Ridolfi_2016, Freire_2017, Kumari_2022}. The black widow PSR J2051$-$0827 exhibits orbital variability characterised by sinusoidal modulation with changing amplitude \citep[see Fig. 5 from][]{Shaifullah_2016}. A chaotic orbital evolution has been also observed in the transitional redback system PSR J1023+0038 during its radio pulsar state \citep{Archibald_2015}, while the orbital period variations seem to be smoother in the X-ray active state \citep{Jaodand_2016, Papitto_2019, Burtovoi_2020, Illiano_2022arXiv}. The long-term orbital modulation of a few black widow pulsars \citep{Arzoumanian_1994, Applegate_Shaham_1994, Doroshenko_2001} has been interpreted in terms of gravitational quadrupole coupling (GQC) model \citep{Applegate_1992, Applegate_Shaham_1994}. This model was  suggested to apply also to the case of SAX J1808 by \citet{Patruno_2012} (see also \citealt{Patruno_2017,Sanna_2017}). It envisages a gravitational coupling between the orbit and variations in the companion quadrupole moment, $\Delta Q$, due to cyclic spin-up and spin-down of the star's outer layers. If $\Delta Q >0$, the companion will become more oblate, its gravitational potential in the equatorial plane will increase, and the orbit will shrink ($\dot{P}_{\textrm{orb}} < 0$). On the contrary, if $\Delta Q < 0$, the companion star will become less oblate, and the orbit will expand ($\dot{P}_{\textrm{orb}} > 0$). Torques produced by magnetic activity of the companion would generate the angular momentum variations that are rapidly transmitted to the orbit.

Given the observed parameters of the long-term oscillation of SAX J1808, the GQC mechanism requires the companion to feature a magnetic field with a strength of $B_2 \simeq 6\times10^{3}$~G and provide an internal luminosity of $L_{\textrm{GQC}}\simeq 10^{30} \, \mathrm{erg \, s^{-1}}$ (see Eqs. (15) and (16) in \citealt{Sanna_2017}, derived from \citealt{Applegate_1992, Applegate_Shaham_1994}), taking the NS and the companion masses equal to $M_{\textrm{NS}} \simeq 1.4\,{M_{\odot}}$ and $M_2 \simeq 0.05\,{M_{\odot}}$, respectively.
However, identifying the energy source required to power such a mechanism in the case of SAX J1808 is not trivial, since nuclear burning or external irradiation of the companion star cannot provide such a high luminosity  \citep{Sanna_2017,Patruno_2017}.

\citet{Sanna_2017} proposed that tidal dissipation could provide such a power if the donor is maintained in asynchronous rotation compared to the orbit by a magnetic braking mechanism. Irradiation by the pulsar wind would sustain the relatively high mass-loss rate needed. In fact, in order to provide the required $L_{\textrm{GQC}}$, the secondary would have to lose mass at a rate
\citep{Applegate_Shaham_1994, Sanna_2017}:
\begin{equation} \label{eq:mass_loss_rate}
    \dot{m}_2 = 0.25 \times 10^{-9} \, \Bigr(\frac{a}{l} \Bigr)^2 \,  L_{30}^{1/2} \, t_{\textrm{sync},4}^{-1/2} \, I_{2,51}^{1/2} \, M_\odot \, \mathrm{yr^{-1}}.
\end{equation}
Here, $a$ is the orbital separation, $l$ is the magnetic lever arm of the mass ejected from the companion star, $L_{30}$ is the tidal luminosity in units of $10^{30} \, \mathrm{erg \, s^{-1}}$,  $I_{2,51}$ is the companion moment of inertia in units of $10^{51} \, \mathrm{g \, cm^2}$, and $t_{\textrm{sync},4}$ is the tidal synchronization time in units of $10^4$ years. For a Roche-lobe filling companion, the latter is estimated as $t_{\textrm{sync}} = 0.65\times10^4 {\mu_{12}}^{-1} (1+q)^2\,  {M_{2,\odot}} {R_{2,\odot}}^{-1}$~yr \citep{Applegate_Shaham_1994}, where $\mu_{12}= 3 \, L_{2,\odot}^{1/3} \, R_{2,\odot}^{-5/3} \, M_{2,\odot}^{2/3}$ is the mean dynamic viscosity in units of $10^{12}\, \mathrm{g \, cm^{-1} \, s^{-1}}$, and $L_{2,\odot}=L_{\textrm{GQC}}/L_\odot$, $M_{2,\odot}$ and $R_{2,\odot}$ are the luminosity, the mass and the radius of the companion star in Solar units, respectively. Assuming $M_2 \approx 0.05 \, {M_\odot}$, $R_2 \approx 0.13 \, {R_\odot}$ \citep{Bildsten_2001}, we obtain $t_{\textrm{sync}} \simeq 3.4 \times 10^3 \, \mathrm{yr}$, similar to the value reported by \citet{Sanna_2017}. 
The corresponding variation of the orbital period in units of $10^{-12} \, \mathrm{s \, s^{-1}}$ is expressed by 
\citep{DiSalvo_2008, Burderi_2009, Sanna_2017}:
\begin{multline} \label{eq:Porb_applegate}
    \dot{P}_{\textrm{orb},-12}=-1.38 \, M_{\textrm{NS},\odot}^{5/3} \, q \, (1+q)^{-1/3} \, P_{\textrm{orb},2h}^{-5/3} \\
    +0.648 \, M_{\textrm{NS},\odot}^{-1} \, q^{-1} \, P_{\textrm{orb},2h} \, g(\beta, q, \alpha) \, \dot{m}_{-9}.
\end{multline} 
Here, $P_{\textrm{orb},2h}$ is the orbital period in units of $2 \, \mathrm{h}$,  $\dot{m}_{-9}=\dot{m}_2/(10^{-9} \, {M_\odot}$\,yr$^{-1})$, $\alpha=\ell_{\textrm{ej}}/\ell_2$ represents the amount of specific angular momentum that is carried away by such an outflow in units of the secondary specific angular momentum, $\beta$ is the fraction of mass lost by the companion that is accreted onto the NS, and $g(\beta, q, \alpha) = 1 - \beta \, q -(1-\beta) \, (\alpha+q/3)/(1+q)$.

Firstly, considering a magnetic lever arm $l\simeq 0.5a$ (similarly to \citealt{Applegate_Shaham_1994, Sanna_2017}), the mass-loss rate is estimated to be $\dot{m}_{-9} \simeq 1.7$ (Eq.~\eqref{eq:mass_loss_rate}). Assuming that only a fraction $\beta=0.01$ of the mass transferred by the companion is accreted by the NS, while the rest is ejected with the specific angular momentum at the inner Lagrangian point ($\alpha=[1-0.462 \, q^{1/3} \, (1+q)^{2/3}]^2 \simeq 0.7$; \citealt{DiSalvo_2008, Burderi_2009, Sanna_2017}) requires a period derivative of $\dot{P}_{\textrm{orb},-12}=7.0$ (Eq.~\eqref{eq:Porb_applegate}). Such a positive derivative seems too large to be compatible with the orbital phase evolution we found. Fixing the $\dot{P}_{\textrm{orb}}$ in Eq.~\eqref{eq:fit_Dtasc} to such a large value and repeating the fit leads to an unreasonably high value of the fit $\chi^2$ ($15817.9/4$). Secondly, assuming a magnetic lever arm $l=a$ (in analogy with what was done in \citealt{Applegate_Shaham_1994}), the mass-loss rate is estimated to be $\dot{m}_{-9} \simeq 0.4$ (Eq.~\eqref{eq:mass_loss_rate}). For $\alpha \simeq 0.7$, we obtain $\dot{P}_{\textrm{orb},-12}=1.6$ (Eq.~\eqref{eq:Porb_applegate}), still too large for the observed orbital evolution ($\chi^2$/dof=$808.5/4$).

By considering a range of orbital period derivative $\dot{P}_{\textrm{orb},-12}$ between 0 and $-0.55$ (i.e., $\pm$ four times the uncertainty on the best-fitting value) we deduce a range of values for $\alpha$  between 1.03 and 1.06 (see Eq.~\eqref{eq:Porb_applegate}) for a mass-loss rate of $\dot{m}_{-9} \simeq 1.7$ ($l \simeq 0.5a$). If we assume $\dot{m}_{-9} \simeq 0.4$ (for $l=a$), the range of value for $\alpha$ is between 1.02 and  1.13.

%A=(\dot{P}_{orb,-12}+1.38 \, M_{NS,\odot}^{5/3} \, q \, (1+q)^{-1/3} \, P_{orb,2h}^{-5/3})
%B=(0.648 \, M_{NS,\odot}^{-1} \, q^{-1} \, P_{orb,2h} \, \dot{m}_{-9})
%g(\beta, q, \alpha) =A/B
%alpha=-(A/B-1+beta *q)/(1-beta)*(1+q)-q/3

While previous models had to assume that mass left the binary system with the specific angular momentum at the inner Lagrangian point (in order to yield a large positive orbital period derivative, see Eq.~\eqref{eq:Porb_applegate}), the analysis of the dataset presented here suggests that the orbit is contracting at a rate one order of magnitude lower than the expansion previously reported. As a consequence, the mass has to leave the system with an angular momentum equal to or greater than that of the secondary center of mass, so as to make the last term in Eq.~\eqref{eq:Porb_applegate} smaller than the first term.
A magnetic slingshot mechanism \citep[see, e.g.,][]{Ferreira_2000, Waugh_2021, Faller_2022} by the strong B-field ($B_2\simeq 6\times10^3$~G) of the companion required to power the observed GQC luminosity might contribute to increase the specific angular momentum carried away by the matter ejected by the pulsar wind from the inner Lagrangian point.
The observations of future outbursts will confirm the parameters of the long-term sinusoidal modulation, and help constrain the sign and magnitude of the orbital period derivative which largely influence the conclusions on the rate of mass loss required to power the GQC mechanism and the location from which mass is ejected.

\section{Conclusions} \label{sec:conclusions}
We presented a coherent timing analysis of \textrm{NICER} observations of SAX J1808.4$-$3658 during its 2022 outburst. 
We updated the orbital solution and investigated the pulse phase evolution during the outburst. We focused on the fundamental frequency, since the second harmonic was often too weak to be detected. We first modelled the phase delays using a constant frequency model, because the addition of a quadratic term (i.e., $\dot{\nu} \neq 0$) did not produce a significant improvement in the data description. Because of the still large phase residuals, we then added to the linear model a dependence of the pulse phase on the flux, following \citet{Bult_2020}, significantly improving the fit's $\chi^2$. We observed an anti-correlation between the phase delays and the source flux, that holds only for count rates lower than $\sim 100$ c/s, i.e., in the reflaring phase.

We confirmed the secular spin-down of $\dot{\nu}_{\textrm{SD}} \simeq -10^{-15}$ \,Hz\,s$^{-1}$, as found in previous works \citep[see e.g.][]{Patruno_2012, Sanna_2017, Bult_2020}, compatible with the energy losses expected from a $\approx 10^{26}$\,G\,cm$^3$ rotating magnetic dipole.

For the first time in the last twenty years, the orbital phase evolution showed evidence that the orbit has contracted since the last epoch. The long-term behaviour of the orbit is described by a $\sim 11$~s modulation with a $\sim 21$~yr period. We excluded the presence of a third body, as the expected Doppler modulation of the pulsar frequency would be about two orders of magnitude higher than observed.

We discussed the observed $\dot{P}_{\textrm{orb}} = - 2.82(69) \times 10^{-13} \, \mathrm{s \, s^{-1}}$ in terms of a coupling between the orbit and variations in the mass quadrupole of the companion star (GQC model; \citealt{Applegate_1992, Applegate_Shaham_1994}). Data suggest that matter leaving the system with the specific angular momentum of the companion center of mass could maintain the donor star out of tidal locking and drive the required oscillation of its structure. A strong magnetisation of the companion star ($B\simeq 6\times10^3$\,G at the surface) is required to couple the mass loss to the donor star's rotation and to increase the angular momentum carried away by the ejected matter compared to the orbital value. 

Based on past recurrence times, it is expected that there will be a new outburst of SAX J1808 in approximately three years (2025). The observations of the next outburst will be of paramount importance to confirm the source’s orbital evolution, by decreasing the correlation between the long-term modulation of the orbital phase epoch and the quadratic term that represents a secular orbital period derivative. This would constrain even more the mass-loss rate and the location from which mass is ejected needed to power the GQC mechanism. 
Detecting pulsations during the quiescent state would greatly increase our ability to track the pulsar orbital evolution without relying solely on data taken during unpredictable outbursts. Even though a rotation-powered pulsar is expected to turn on during quiescence \citep{Burderi_2003}, deep searches for radio \citep{Burgay_2003, Patruno_2017} or gamma-ray \citep{2016MNRAS.456.2647D} pulsations have not succeeded in detecting a signal, so far. Recently, the discovery of optical pulsations from a couple of millisecond pulsars (\citealt{Ambrosino_2017, Ambrosino_2021}; Miraval Zanon et al. in prep) opened the intriguing possibility of exploiting the higher sensitivity in this band compared to higher energies, and we plan to use this additional diagnostic also to investigate the orbital evolution of this source.\\ \\
This work is based on observations acquired with the NASA mission \textrm{NICER}. This research has made use of data and software provided by the High Energy Astrophysics Science Archive Research Center (HEASARC), which is a service of the Astrophysics Science Division at NASA/GSFC.
G.I. is supported by the AASS Ph.D. joint research program between the University of Rome ``Sapienza'' and the University of Rome ``Tor Vergata'', with the collaboration of the National Institute of Astrophysics (INAF). F.A., G.I., A.M.Z., A.P., L.S., and D.d.M. acknowledge financial support from the Italian Space Agency (ASI) and National Institute for Astrophysics (INAF) under agreements ASI-INAF I/037/12/0 and ASI-INAF n.2017-14-H.0, from INAF Research Grant ``Uncovering the optical beat of the fastest magnetised neutron stars (FANS)''.
F.A., G.I., A.M.Z., A.P., and L.S. also acknowledge funding from the Italian Ministry of University and Research (MUR), PRIN 2020 (prot. 2020BRP57Z) ``Gravitational and Electromagnetic-wave Sources in the Universe with current and next-generation detectors (GEMS)''. L.S. and T.D.S. acknowledge financial contributions from ``iPeska'' research grant (P.I. Andrea Possenti) funded under the INAF call PRIN-SKA/CTA (resolution 70/2016).
L.S. is also partially supported by PRIN-INAF 2019 no. 15. T.D.S. acknowledges financial support from PRIN-INAF 2019 (n. 89, PI: Belloni). A.M.Z. is supported by PRIN-MIUR 2017 UnIAM (Unifying Isolated and Accreting Magnetars, PI S. Mereghetti). P.B. acknowledges support from NASA through CRESST II cooperative agreement (80GSFC21M0002). F.C.Z. and A.M. are supported by the H2020 ERC Consolidator Grant ``MAGNESIA'' under grant agreement No. 817661 (PI: Rea) and National Spanish grant PGC2018-095512-BI00. This work was also partially supported by the program Unidad de Excelencia Maria de Maeztu CEX2020-001058-M, and by the PHAROS COST Action (No. CA16214). F.C.Z. is also supported by Juan de la Cierva Fellowship. 
J.P. acknowledges support from the Academy of Finland grant 333112.


\facilities{ADS, HEASARC, \textrm{NICER}}
\software{HEASoft (v6.30), NICERDAS (v7a), matplotlib \citep{Hunter_2007}, scipy \citep{Virtanen_2020}}

\bibliography{biblio}{}

\begin{thebibliography}{}
\expandafter\ifx\csname natexlab\endcsname\relax\def\natexlab#1{#1}\fi
\providecommand{\url}[1]{\href{#1}{#1}}
\providecommand{\dodoi}[1]{doi:~\href{http://doi.org/#1}{\nolinkurl{#1}}}
\providecommand{\doeprint}[1]{\href{http://ascl.net/#1}{\nolinkurl{http://ascl.net/#1}}}
\providecommand{\doarXiv}[1]{\href{https://arxiv.org/abs/#1}{\nolinkurl{https://arxiv.org/abs/#1}}}

\bibitem[{{Ambrosino} {et~al.}(2017){Ambrosino}, {Papitto}, {Stella}, {Meddi},
  {Cretaro}, {Burderi}, {Di Salvo}, {Israel}, {Ghedina}, {Di Fabrizio}, \&
  {Riverol}}]{Ambrosino_2017}
{Ambrosino}, F., {Papitto}, A., {Stella}, L., {et~al.} 2017, Nature Astronomy,
  1, 854, \dodoi{10.1038/s41550-017-0266-2}

\bibitem[{{Ambrosino} {et~al.}(2021){Ambrosino}, {Miraval Zanon}, {Papitto},
  {Coti Zelati}, {Campana}, {D'Avanzo}, {Stella}, {Di Salvo}, {Burderi},
  {Casella}, {Sanna}, {de Martino}, {Cadelano}, {Ghedina}, {Leone}, {Meddi},
  {Cretaro}, {Baglio}, {Poretti}, {Mignani}, {Torres}, {Israel}, {Cecconi},
  {Russell}, {Gonzalez Gomez}, {Riverol Rodriguez}, {Perez Ventura}, {Hernandez
  Diaz}, {San Juan}, {Bramich}, \& {Lewis}}]{Ambrosino_2021}
{Ambrosino}, F., {Miraval Zanon}, A., {Papitto}, A., {et~al.} 2021, Nature
  Astronomy, 5, 552, \dodoi{10.1038/s41550-021-01308-0}

\bibitem[{{Applegate}(1992)}]{Applegate_1992}
{Applegate}, J.~H. 1992, \apj, 385, 621, \dodoi{10.1086/170967}

\bibitem[{{Applegate} \& {Shaham}(1994)}]{Applegate_Shaham_1994}
{Applegate}, J.~H., \& {Shaham}, J. 1994, \apj, 436, 312,
  \dodoi{10.1086/174906}

\bibitem[{{Archibald} {et~al.}(2015){Archibald}, {Bogdanov}, {Patruno},
  {Hessels}, {Deller}, {Bassa}, {Janssen}, {Kaspi}, {Lyne}, {Stappers},
  {Tendulkar}, {D'Angelo}, \& {Wijnands}}]{Archibald_2015}
{Archibald}, A.~M., {Bogdanov}, S., {Patruno}, A., {et~al.} 2015, \apj, 807,
  62, \dodoi{10.1088/0004-637X/807/1/62}

\bibitem[{{Arnaud}(1996)}]{Arnaud_1996}
{Arnaud}, K.~A. 1996, in Astronomical Society of the Pacific Conference Series,
  Vol. 101, Astronomical Data Analysis Software and Systems V, ed. G.~H.
  {Jacoby} \& J.~{Barnes}, 17

\bibitem[{{Arzoumanian} {et~al.}(1994){Arzoumanian}, {Fruchter}, \&
  {Taylor}}]{Arzoumanian_1994}
{Arzoumanian}, Z., {Fruchter}, A.~S., \& {Taylor}, J.~H. 1994, \apjl, 426, L85,
  \dodoi{10.1086/187346}

\bibitem[{{Avni}(1976)}]{Avni_1976}
{Avni}, Y. 1976, \apj, 210, 642, \dodoi{10.1086/154870}

\bibitem[{{Bildsten} \& {Chakrabarty}(2001)}]{Bildsten_2001}
{Bildsten}, L., \& {Chakrabarty}, D. 2001, \apj, 557, 292,
  \dodoi{10.1086/321633}

\bibitem[{{Bult} {et~al.}(2020){Bult}, {Chakrabarty}, {Arzoumanian},
  {Gendreau}, {Guillot}, {Malacaria}, {Ray}, \& {Strohmayer}}]{Bult_2020}
{Bult}, P., {Chakrabarty}, D., {Arzoumanian}, Z., {et~al.} 2020, \apj, 898, 38,
  \dodoi{10.3847/1538-4357/ab9827}

\bibitem[{{Bult} \& {van der Klis}(2015)}]{Bult_2015}
{Bult}, P., \& {van der Klis}, M. 2015, \apj, 806, 90,
  \dodoi{10.1088/0004-637X/806/1/90}

\bibitem[{{Bult} {et~al.}(2022){Bult}, {Altamirano}, {Arzoumanian},
  {Chakrabarty}, {Chenevez}, {Ferrara}, {Gendreau}, {Guillot}, {G{\"u}ver},
  {Iwakiri}, {Jaisawal}, {Mancuso}, {Malacaria}, {Ng}, {Sanna}, {Strohmayer},
  {Wadiasingh}, \& {Wolff}}]{Bult_2022}
{Bult}, P., {Altamirano}, D., {Arzoumanian}, Z., {et~al.} 2022, \apjl, 935,
  L32, \dodoi{10.3847/2041-8213/ac87f9}

\bibitem[{{Burderi} {et~al.}(2003){Burderi}, {Di Salvo}, {D'Antona}, {Robba},
  \& {Testa}}]{Burderi_2003}
{Burderi}, L., {Di Salvo}, T., {D'Antona}, F., {Robba}, N.~R., \& {Testa}, V.
  2003, \aap, 404, L43, \dodoi{10.1051/0004-6361:20030669}

\bibitem[{{Burderi} {et~al.}(2006){Burderi}, {Di Salvo}, {Menna}, {Riggio}, \&
  {Papitto}}]{Burderi_2006}
{Burderi}, L., {Di Salvo}, T., {Menna}, M.~T., {Riggio}, A., \& {Papitto}, A.
  2006, \apjl, 653, L133, \dodoi{10.1086/510666}

\bibitem[{{Burderi} {et~al.}(2009){Burderi}, {Riggio}, {Di Salvo}, {Papitto},
  {Menna}, {D'A{\`\i}}, \& {Iaria}}]{Burderi_2009}
{Burderi}, L., {Riggio}, A., {Di Salvo}, T., {et~al.} 2009, \aap, 496, L17,
  \dodoi{10.1051/0004-6361/200811542}

\bibitem[{{Burderi} {et~al.}(2007){Burderi}, {Di Salvo}, {Lavagetto}, {Menna},
  {Papitto}, {Riggio}, {Iaria}, {D'Antona}, {Robba}, \&
  {Stella}}]{Burderi_2007}
{Burderi}, L., {Di Salvo}, T., {Lavagetto}, G., {et~al.} 2007, \apj, 657, 961,
  \dodoi{10.1086/510659}

\bibitem[{{Burgay} {et~al.}(2003){Burgay}, {Burderi}, {Possenti}, {D'Amico},
  {Manchester}, {Lyne}, {Camilo}, \& {Campana}}]{Burgay_2003}
{Burgay}, M., {Burderi}, L., {Possenti}, A., {et~al.} 2003, \apj, 589, 902,
  \dodoi{10.1086/374690}

\bibitem[{{Burtovoi} {et~al.}(2020){Burtovoi}, {Zampieri}, {Fiori}, {Naletto},
  {Spolon}, {Barbieri}, {Papitto}, \& {Ambrosino}}]{Burtovoi_2020}
{Burtovoi}, A., {Zampieri}, L., {Fiori}, M., {et~al.} 2020, \mnras, 498, L98,
  \dodoi{10.1093/mnrasl/slaa133}

\bibitem[{{Campana} {et~al.}(2004){Campana}, {D'Avanzo}, {Casares}, {Covino},
  {Israel}, {Marconi}, {Hynes}, {Charles}, \& {Stella}}]{Campana_2004}
{Campana}, S., {D'Avanzo}, P., {Casares}, J., {et~al.} 2004, \apjl, 614, L49,
  \dodoi{10.1086/425495}

\bibitem[{{Chakrabarty} \& {Morgan}(1998)}]{Chakrabarty_Morgan_1998}
{Chakrabarty}, D., \& {Morgan}, E.~H. 1998, \nat, 394, 346,
  \dodoi{10.1038/28561}

\bibitem[{{Chakrabarty} {et~al.}(2003){Chakrabarty}, {Morgan}, {Muno},
  {Galloway}, {Wijnands}, {van der Klis}, \& {Markwardt}}]{Chakrabarty_2003}
{Chakrabarty}, D., {Morgan}, E.~H., {Muno}, M.~P., {et~al.} 2003, \nat, 424,
  42, \dodoi{10.1038/nature01732}

\bibitem[{{Cui} {et~al.}(1998){Cui}, {Morgan}, \& {Titarchuk}}]{Cui_1998}
{Cui}, W., {Morgan}, E.~H., \& {Titarchuk}, L.~G. 1998, \apjl, 504, L27,
  \dodoi{10.1086/311569}

\bibitem[{{de O{\~n}a Wilhelmi} {et~al.}(2016){de O{\~n}a Wilhelmi}, {Papitto},
  {Li}, {Rea}, {Torres}, {Burderi}, {Di Salvo}, {Iaria}, {Riggio}, \&
  {Sanna}}]{2016MNRAS.456.2647D}
{de O{\~n}a Wilhelmi}, E., {Papitto}, A., {Li}, J., {et~al.} 2016, \mnras, 456,
  2647, \dodoi{10.1093/mnras/stv2695}

\bibitem[{{Deeter} {et~al.}(1981){Deeter}, {Boynton}, \&
  {Pravdo}}]{Deeter_Boyton_Pravdo_1981}
{Deeter}, J.~E., {Boynton}, P.~E., \& {Pravdo}, S.~H. 1981, \apj, 247, 1003,
  \dodoi{10.1086/159110}

\bibitem[{{Di Salvo} {et~al.}(2008){Di Salvo}, {Burderi}, {Riggio}, {Papitto},
  \& {Menna}}]{DiSalvo_2008}
{Di Salvo}, T., {Burderi}, L., {Riggio}, A., {Papitto}, A., \& {Menna}, M.~T.
  2008, \mnras, 389, 1851, \dodoi{10.1111/j.1365-2966.2008.13709.x}

\bibitem[{{Di Salvo} \& {Sanna}(2022)}]{DiSalvo_2022}
{Di Salvo}, T., \& {Sanna}, A. 2022, in Astrophysics and Space Science Library,
  Vol. 465, Astrophysics and Space Science Library, ed. S.~{Bhattacharyya},
  A.~{Papitto}, \& D.~{Bhattacharya}, 87--124,
  \dodoi{10.1007/978-3-030-85198-9_4}

\bibitem[{{Di Salvo} {et~al.}(2019){Di Salvo}, {Sanna}, {Burderi}, {Papitto},
  {Iaria}, {Gambino}, \& {Riggio}}]{DiSalvo_2019}
{Di Salvo}, T., {Sanna}, A., {Burderi}, L., {et~al.} 2019, \mnras, 483, 767,
  \dodoi{10.1093/mnras/sty2974}

\bibitem[{{Doroshenko} {et~al.}(2001){Doroshenko}, {L{\"o}hmer}, {Kramer},
  {Jessner}, {Wielebinski}, {Lyne}, \& {Lange}}]{Doroshenko_2001}
{Doroshenko}, O., {L{\"o}hmer}, O., {Kramer}, M., {et~al.} 2001, \aap, 379,
  579, \dodoi{10.1051/0004-6361:20011349}

\bibitem[{{Faller} \& {Jardine}(2022)}]{Faller_2022}
{Faller}, S.~J., \& {Jardine}, M.~M. 2022, \mnras, 513, 5611,
  \dodoi{10.1093/mnras/stac1273}

\bibitem[{{Ferreira}(2000)}]{Ferreira_2000}
{Ferreira}, J.~M. 2000, \mnras, 316, 647,
  \dodoi{10.1046/j.1365-8711.2000.03540.x}

\bibitem[{{Finger} {et~al.}(1999){Finger}, {Bildsten}, {Chakrabarty}, {Prince},
  {Scott}, {Wilson}, {Wilson}, \& {Zhang}}]{Finger_1999}
{Finger}, M.~H., {Bildsten}, L., {Chakrabarty}, D., {et~al.} 1999, \apj, 517,
  449, \dodoi{10.1086/307191}

\bibitem[{{Freire} {et~al.}(2017){Freire}, {Ridolfi}, {Kramer}, {Jordan},
  {Manchester}, {Torne}, {Sarkissian}, {Heinke}, {D'Amico}, {Camilo},
  {Lorimer}, \& {Lyne}}]{Freire_2017}
{Freire}, P.~C.~C., {Ridolfi}, A., {Kramer}, M., {et~al.} 2017, \mnras, 471,
  857, \dodoi{10.1093/mnras/stx1533}

\bibitem[{{Galloway} \& {Cumming}(2006)}]{Galloway_2006}
{Galloway}, D.~K., \& {Cumming}, A. 2006, \apj, 652, 559,
  \dodoi{10.1086/507598}

\bibitem[{{Gendreau} {et~al.}(2012){Gendreau}, {Arzoumanian}, \&
  {Okajima}}]{NICER_Gendreau_2012}
{Gendreau}, K.~C., {Arzoumanian}, Z., \& {Okajima}, T. 2012, in Society of
  Photo-Optical Instrumentation Engineers (SPIE) Conference Series, Vol. 8443,
  Space Telescopes and Instrumentation 2012: Ultraviolet to Gamma Ray, ed.
  T.~{Takahashi}, S.~S. {Murray}, \& J.-W.~A. {den Herder}, 844313,
  \dodoi{10.1117/12.926396}

\bibitem[{{Gilfanov} {et~al.}(1998){Gilfanov}, {Revnivtsev}, {Sunyaev}, \&
  {Churazov}}]{Gilfanov_1998}
{Gilfanov}, M., {Revnivtsev}, M., {Sunyaev}, R., \& {Churazov}, E. 1998, \aap,
  338, L83.
\newblock \doarXiv{astro-ph/9805152}

\bibitem[{{Goodwin} {et~al.}(2019){Goodwin}, {Galloway}, {Heger}, {Cumming}, \&
  {Johnston}}]{Goodwin_2019}
{Goodwin}, A.~J., {Galloway}, D.~K., {Heger}, A., {Cumming}, A., \& {Johnston},
  Z. 2019, \mnras, 490, 2228, \dodoi{10.1093/mnras/stz2638}

\bibitem[{{Hartman} {et~al.}(2009){Hartman}, {Patruno}, {Chakrabarty},
  {Markwardt}, {Morgan}, {van der Klis}, \& {Wijnands}}]{Hartman_2009}
{Hartman}, J.~M., {Patruno}, A., {Chakrabarty}, D., {et~al.} 2009, \apj, 702,
  1673, \dodoi{10.1088/0004-637X/702/2/1673}

\bibitem[{{Hartman} {et~al.}(2008){Hartman}, {Patruno}, {Chakrabarty},
  {Kaplan}, {Markwardt}, {Morgan}, {Ray}, {van der Klis}, \&
  {Wijnands}}]{Hartman_2008}
---. 2008, \apj, 675, 1468, \dodoi{10.1086/527461}

\bibitem[{Hunter(2007)}]{Hunter_2007}
Hunter, J.~D. 2007, Computing in Science \& Engineering, 9, 90,
  \dodoi{10.1109/MCSE.2007.55}

\bibitem[{{Illiano} {et~al.}(2022{\natexlab{a}}){Illiano}, {Papitto}, {Zanon},
  {Ambrosino}, {Baglio}, {Sanna}, {Burderi}, {Bult}, {Ng}, {Chakrabarty}, {Di
  Salvo}, \& {Altamirano}}]{Illiano_2022ATel}
{Illiano}, G., {Papitto}, A., {Zanon}, A.~M., {et~al.} 2022{\natexlab{a}}, The
  Astronomer's Telegram, 15647, 1

\bibitem[{{Illiano} {et~al.}(2022{\natexlab{b}}){Illiano}, {Papitto},
  {Ambrosino}, {Miraval Zanon}, {Coti Zelati}, {Stella}, {Zampieri},
  {Burtovoi}, {Campana}, {Casella}, {Cecconi}, {de Martino}, {Fiori},
  {Ghedina}, {Gonzales}, {Hernandez Diaz}, {Israel}, {Leone}, {Naletto}, {Perez
  Ventura}, {Riverol}, {Riverol}, {Torres}, \& {Turchetta}}]{Illiano_2022arXiv}
{Illiano}, G., {Papitto}, A., {Ambrosino}, F., {et~al.} 2022{\natexlab{b}},
  arXiv e-prints, arXiv:2211.12975.
\newblock \doarXiv{2211.12975}

\bibitem[{{Imai} {et~al.}(2022){Imai}, {Serino}, {Negoro}, {Nakajima},
  {Kobayashi}, {Tanaka}, {Soejima}, {Mihara}, {Kawamuro}, {Yamada}, {Tamagawa},
  {Matsuoka}, {Sakamoto}, {Sugita}, {Hiramatsu}, {Yoshida}, {Tsuboi},
  {Iwakiri}, {Kohara}, {Shidatsu}, {Iwasaki}, {Kawai}, {Niwano}, {Hosokawa},
  {Ito}, {Takamatsu}, {Nakahira}, {Ueno}, {Tomida}, {Ishikawa}, {Kurihara},
  {Ueda}, {Ogawa}, {Setoguchi}, {Yoshitake}, {Inaba}, {Yamauchi}, {Sato},
  {Hatsuda}, {Fukuoka}, {Hagiwara}, {Umeki}, {Yamaoka}, {Kawakubo}, \&
  {Sugizaki}}]{Atel_MAXI_2022}
{Imai}, Y., {Serino}, M., {Negoro}, H., {et~al.} 2022, The Astronomer's
  Telegram, 15563, 1

\bibitem[{{in 't Zand} {et~al.}(1998){in 't Zand}, {Heise}, {Muller},
  {Bazzano}, {Cocchi}, {Natalucci}, \& {Ubertini}}]{Discovery_1996}
{in 't Zand}, J.~J.~M., {Heise}, J., {Muller}, J.~M., {et~al.} 1998, \aap, 331,
  L25.
\newblock \doarXiv{astro-ph/9802098}

\bibitem[{{in't Zand} {et~al.}(2001){in't Zand}, {Cornelisse}, {Kuulkers},
  {Heise}, {Kuiper}, {Bazzano}, {Cocchi}, {Muller}, {Natalucci}, {Smith}, \&
  {Ubertini}}]{intZand_2001}
{in't Zand}, J.~J.~M., {Cornelisse}, R., {Kuulkers}, E., {et~al.} 2001, \aap,
  372, 916, \dodoi{10.1051/0004-6361:20010546}

\bibitem[{{Jaodand} {et~al.}(2016){Jaodand}, {Archibald}, {Hessels},
  {Bogdanov}, {D'Angelo}, {Patruno}, {Bassa}, \& {Deller}}]{Jaodand_2016}
{Jaodand}, A., {Archibald}, A.~M., {Hessels}, J. W.~T., {et~al.} 2016, \apj,
  830, 122, \dodoi{10.3847/0004-637X/830/2/122}

\bibitem[{{Kulkarni} \& {Romanova}(2013)}]{Kulkarni_2013}
{Kulkarni}, A.~K., \& {Romanova}, M.~M. 2013, \mnras, 433, 3048,
  \dodoi{10.1093/mnras/stt945}

\bibitem[{{Kumari} {et~al.}(2022){Kumari}, {Bhattacharyya}, {Kansabanik}, \&
  {Roy}}]{Kumari_2022}
{Kumari}, S., {Bhattacharyya}, B., {Kansabanik}, D., \& {Roy}, J. 2022, arXiv
  e-prints, arXiv:2211.14107.
\newblock \doarXiv{2211.14107}

\bibitem[{{Lampton} {et~al.}(1976){Lampton}, {Margon}, \&
  {Bowyer}}]{Lampton_Margon_Bowyer}
{Lampton}, M., {Margon}, B., \& {Bowyer}, S. 1976, \apj, 208, 177,
  \dodoi{10.1086/154592}

\bibitem[{{Lyne} \& {Graham-Smith}(1990)}]{Lyne_GrahamSmith_1990}
{Lyne}, A.~G., \& {Graham-Smith}, F. 1990, Cambridge Astrophysics Series, 16

\bibitem[{{Marino} {et~al.}(2019){Marino}, {Di Salvo}, {Burderi}, {Sanna},
  {Riggio}, {Papitto}, {Del Santo}, {Gambino}, {Iaria}, \&
  {Mazzola}}]{Marino_2019}
{Marino}, A., {Di Salvo}, T., {Burderi}, L., {et~al.} 2019, \aap, 627, A125,
  \dodoi{10.1051/0004-6361/201834460}

\bibitem[{{Papitto} {et~al.}(2007){Papitto}, {Di Salvo}, {Burderi}, {Menna},
  {Lavagetto}, \& {Riggio}}]{Papitto_2007}
{Papitto}, A., {Di Salvo}, T., {Burderi}, L., {et~al.} 2007, \mnras, 375, 971,
  \dodoi{10.1111/j.1365-2966.2006.11359.x}

\bibitem[{{Papitto} {et~al.}(2011){Papitto}, {Riggio}, {Burderi}, {Di Salvo},
  {D'Ai'}, {Iaria}, {Bozzo}, {Ferrigno}, \& {Belloni}}]{Atel_Papitto_2011}
{Papitto}, A., {Riggio}, A., {Burderi}, L., {et~al.} 2011, The Astronomer's
  Telegram, 3757, 1

\bibitem[{{Papitto} {et~al.}(2019){Papitto}, {Ambrosino}, {Stella}, {Torres},
  {Coti Zelati}, {Ghedina}, {Meddi}, {Sanna}, {Casella}, {Dallilar},
  {Eikenberry}, {Israel}, {Onori}, {Piranomonte}, {Bozzo}, {Burderi},
  {Campana}, {de Martino}, {Di Salvo}, {Ferrigno}, {Rea}, {Riggio}, {Serrano},
  {Veledina}, \& {Zampieri}}]{Papitto_2019}
{Papitto}, A., {Ambrosino}, F., {Stella}, L., {et~al.} 2019, \apj, 882, 104,
  \dodoi{10.3847/1538-4357/ab2fdf}

\bibitem[{{Patruno} {et~al.}(2012){Patruno}, {Bult}, {Gopakumar}, {Hartman},
  {Wijnands}, {van der Klis}, \& {Chakrabarty}}]{Patruno_2012}
{Patruno}, A., {Bult}, P., {Gopakumar}, A., {et~al.} 2012, \apjl, 746, L27,
  \dodoi{10.1088/2041-8205/746/2/L27}

\bibitem[{{Patruno} {et~al.}(2016){Patruno}, {Maitra}, {Curran}, {D'Angelo},
  {Fridriksson}, {Russell}, {Middleton}, \& {Wijnands}}]{Patruno_2016}
{Patruno}, A., {Maitra}, D., {Curran}, P.~A., {et~al.} 2016, \apj, 817, 100,
  \dodoi{10.3847/0004-637X/817/2/100}

\bibitem[{{Patruno} {et~al.}(2009){Patruno}, {Rea}, {Altamirano}, {Linares},
  {Wijnands}, \& {van der Klis}}]{Patruno_2009}
{Patruno}, A., {Rea}, N., {Altamirano}, D., {et~al.} 2009, \mnras, 396, L51,
  \dodoi{10.1111/j.1745-3933.2009.00660.x}

\bibitem[{{Patruno} \& {Watts}(2021)}]{Patruno_2021}
{Patruno}, A., \& {Watts}, A.~L. 2021, in Astrophysics and Space Science
  Library, Vol. 461, Astrophysics and Space Science Library, ed. T.~M.
  {Belloni}, M.~{M{\'e}ndez}, \& C.~{Zhang}, 143--208,
  \dodoi{10.1007/978-3-662-62110-3_4}

\bibitem[{{Patruno} {et~al.}(2017){Patruno}, {Jaodand}, {Kuiper}, {Bult},
  {Hessels}, {Knigge}, {King}, {Wijnands}, \& {van der Klis}}]{Patruno_2017}
{Patruno}, A., {Jaodand}, A., {Kuiper}, L., {et~al.} 2017, \apj, 841, 98,
  \dodoi{10.3847/1538-4357/aa6f5b}

\bibitem[{{Remillard} {et~al.}(2022){Remillard}, {Loewenstein}, {Steiner},
  {Prigozhin}, {LaMarr}, {Enoto}, {Gendreau}, {Arzoumanian}, {Markwardt},
  {Basak}, {Stevens}, {Ray}, {Altamirano}, \& {Buisson}}]{Remillard_2022}
{Remillard}, R.~A., {Loewenstein}, M., {Steiner}, J.~F., {et~al.} 2022, \aj,
  163, 130, \dodoi{10.3847/1538-3881/ac4ae6}

\bibitem[{{Ridolfi} {et~al.}(2016){Ridolfi}, {Freire}, {Torne}, {Heinke}, {van
  den Berg}, {Jordan}, {Kramer}, {Bassa}, {Sarkissian}, {D'Amico}, {Lorimer},
  {Camilo}, {Manchester}, \& {Lyne}}]{Ridolfi_2016}
{Ridolfi}, A., {Freire}, P.~C.~C., {Torne}, P., {et~al.} 2016, \mnras, 462,
  2918, \dodoi{10.1093/mnras/stw1850}

\bibitem[{{Sanna} {et~al.}(2017){Sanna}, {Di Salvo}, {Burderi}, {Riggio},
  {Pintore}, {Gambino}, {Iaria}, {Tailo}, {Scarano}, \& {Papitto}}]{Sanna_2017}
{Sanna}, A., {Di Salvo}, T., {Burderi}, L., {et~al.} 2017, \mnras, 471, 463,
  \dodoi{10.1093/mnras/stx1588}

\bibitem[{{Sanna} {et~al.}(2022{\natexlab{a}}){Sanna}, {Bult}, {Gendreau},
  {Arzoumanian}, {Jaisawal}, {Altamirano}, {Ng}, {Chakrabarty}, {Ray},
  {Negoro}, {Iwakiri}, {Mihara}, \& {Serino}}]{Atel_Sanna_2022}
{Sanna}, A., {Bult}, P.~M., {Gendreau}, K.~C., {et~al.} 2022{\natexlab{a}}, The
  Astronomer's Telegram, 15559, 1

\bibitem[{{Sanna} {et~al.}(2022{\natexlab{b}}){Sanna}, {Burderi}, {Di Salvo},
  {Riggio}, {Altamirano}, {Marino}, {Bult}, {Strohmayer}, {Guillot},
  {Malacaria}, {Ng}, {Mancuso}, {Mazzola}, {Albayati}, {Iaria}, {Manca},
  {Deiosso}, {Cabras}, \& {Anitra}}]{Sanna_2022}
{Sanna}, A., {Burderi}, L., {Di Salvo}, T., {et~al.} 2022{\natexlab{b}},
  \mnras, 514, 4385, \dodoi{10.1093/mnras/stac1611}

\bibitem[{{Shaifullah} {et~al.}(2016){Shaifullah}, {Verbiest}, {Freire},
  {Tauris}, {Wex}, {Os{\l}owski}, {Stappers}, {Bassa}, {Caballero}, {Champion},
  {Cognard}, {Desvignes}, {Graikou}, {Guillemot}, {Janssen}, {Jessner},
  {Jordan}, {Karuppusamy}, {Kramer}, {Lazaridis}, {Lazarus}, {Lyne}, {McKee},
  {Perrodin}, {Possenti}, \& {Tiburzi}}]{Shaifullah_2016}
{Shaifullah}, G., {Verbiest}, J.~P.~W., {Freire}, P.~C.~C., {et~al.} 2016,
  \mnras, 462, 1029, \dodoi{10.1093/mnras/stw1737}

\bibitem[{{Standish}(1998)}]{Standish_DE405}
{Standish}, E.~M. 1998, JPL Planetary and Lunar Ephemerides, DE405/LE405, JPL
  Interoffice Memo 312.F-98-048 (Pasadena, CA: NASA Jet Propulsion Laboratory)

\bibitem[{{Stella} {et~al.}(2000){Stella}, {Campana}, {Mereghetti}, {Ricci}, \&
  {Israel}}]{Stella_2000}
{Stella}, L., {Campana}, S., {Mereghetti}, S., {Ricci}, D., \& {Israel}, G.~L.
  2000, \apjl, 537, L115, \dodoi{10.1086/312765}

\bibitem[{{Verner} {et~al.}(1996){Verner}, {Ferland}, {Korista}, \&
  {Yakovlev}}]{Verner_1996}
{Verner}, D.~A., {Ferland}, G.~J., {Korista}, K.~T., \& {Yakovlev}, D.~G. 1996,
  \apj, 465, 487, \dodoi{10.1086/177435}

\bibitem[{{Virtanen} {et~al.}(2020){Virtanen}, {Gommers}, {Oliphant},
  {Haberland}, {Reddy}, {Cournapeau}, {Burovski}, {Peterson}, {Weckesser},
  {Bright}, {van der Walt}, {Brett}, {Wilson}, {Millman}, {Mayorov}, {Nelson},
  {Jones}, {Kern}, {Larson}, {Carey}, {Polat}, {Feng}, {Moore}, {VanderPlas},
  {Laxalde}, {Perktold}, {Cimrman}, {Henriksen}, {Quintero}, {Harris},
  {Archibald}, {Ribeiro}, {Pedregosa}, {van Mulbregt}, \& {SciPy 1. 0
  Contributors}}]{Virtanen_2020}
{Virtanen}, P., {Gommers}, R., {Oliphant}, T.~E., {et~al.} 2020, Nature
  Methods, 17, 261, \dodoi{10.1038/s41592-019-0686-2}

\bibitem[{{Waugh} {et~al.}(2021){Waugh}, {Jardine}, {Morin}, \&
  {Donati}}]{Waugh_2021}
{Waugh}, R. F.~P., {Jardine}, M.~M., {Morin}, J., \& {Donati}, J.~F. 2021,
  \mnras, 505, 5104, \dodoi{10.1093/mnras/stab1709}

\bibitem[{{Wijnands} {et~al.}(2001){Wijnands}, {M{\'e}ndez}, {Markwardt}, {van
  der Klis}, {Chakrabarty}, \& {Morgan}}]{Wijnands_2001}
{Wijnands}, R., {M{\'e}ndez}, M., {Markwardt}, C., {et~al.} 2001, \apj, 560,
  892, \dodoi{10.1086/323073}

\bibitem[{{Wijnands} \& {van der Klis}(1998)}]{Wijnands_1998}
{Wijnands}, R., \& {van der Klis}, M. 1998, \nat, 394, 344,
  \dodoi{10.1038/28557}

\bibitem[{{Wilms} {et~al.}(2000){Wilms}, {Allen}, \& {McCray}}]{Wilms_2000}
{Wilms}, J., {Allen}, A., \& {McCray}, R. 2000, \apj, 542, 914,
  \dodoi{10.1086/317016}

\bibitem[{{Yaqoob}(1998)}]{Yaqoob_1998}
{Yaqoob}, T. 1998, \apj, 500, 893, \dodoi{10.1086/305781}

\end{thebibliography}
\bibliographystyle{aasjournal}

\end{document}